\documentclass{IEEEtran}
\usepackage{amsmath,amssymb,amsfonts}
\usepackage{graphicx}
\usepackage{textcomp,nicefrac}

\usepackage{hyperref}
\usepackage{textcomp, gensymb}
\usepackage{xcolor}
\usepackage{sidecap}
\usepackage{soul}
\usepackage{dirtytalk}
\usepackage{float}
\usepackage{subfigure}
\usepackage{algorithm}
\usepackage{algpseudocode}

\begin{document}
\title{Simulations of Sparse Static Detector Networks for City-Scale Radiological/Nuclear Detection}
\author{E. Rofors, N. Abgrall, M.S. Bandstra, R.J. Cooper, D. Hellfeld, T.H.Y. Joshi, V. Negut, B.J. Quiter, and M. Salathe
\thanks{This work was performed under the auspices of the U.S. Department of Energy by Lawrence Berkeley National Laboratory (LBNL) under Contract DE-AC02-05CH11231.  The project was funded by the U.S. Department of Energy, National Nuclear Security Administration, Office of Defense Nuclear Nonproliferation Research and Development. At the time of this work, all authors were at the Lawrence Berkeley National Laboratory, Berkeley, CA 94720, USA. E-mail: erofors@lbl.gov}}

\maketitle





\begin{abstract}
Sparse static detector networks in urban environments can be used in efforts to detect illicit radioactive sources, such as stolen nuclear material or radioactive \say{dirty bombs}.
We use detailed simulations to evaluate multiple configurations of detector networks and their ability to detect sources moving through a $6\times6$\,km$^2$ area of downtown Chicago. A detector network's probability of detecting a source increases with detector density but can also be increased with strategic node placement. We show that the ability to fuse correlated data from a source-carrying vehicle passing by multiple detectors can significantly contribute to the overall detection probability. In this paper we distinguish static sensor deployments operated as networks able to correlate signals between sensors, from deployments operated as arrays where each sensor is operated individually. In particular, we show that additional visual attributes of source-carrying vehicles, such as vehicle color and make, can greatly improve the ability of a detector network to detect illicit sources.
\end{abstract}


\section{Introduction}
\label{intro}
Risks associated with illicit activities such as smuggling of nuclear materials or deployment of \say{dirty bombs} in urban environments underscore the need for efficient and reliable radiation detection systems. Over the years, numerous approaches centered around mobile and/or stationary radiation detectors have been suggested to counter these threats. Mobile detectors mounted on vehicles have been suggested as able to cover wide areas and actively respond to available threat information~\cite{king2010,hochbaum2011,brown2023,schmidt2019,Flanagan2021}. Detectors installed at fixed locations provide continuous coverage of points of interest and, with enough nodes, can monitor entire areas. Static detector networks have been studied using both simulated and measured data~\cite{nemzek2004, chandy2008, Flanagan2024}. Increasing the number of detectors in a network is a straight-forward approach to enhance the probability of detecting illicit sources. However, cost is a limiting factor for the number of nodes that can be realistically deployed in a city~\cite{wein2007}. One promising approach to cover wider areas with a limited number of detectors is to use sparse static detector networks. This work is performed in the context of one such sparse static detector network that is being developed through the Platforms and Algorithms for Networked Detection and Analysis (PANDA) and Domain Aware Waggle Network (DAWN) projects. These projects are led by Lawrence Berkeley National Laboratory and Argonne National Laboratory, respectively, and collectively referred to as PANDAWN~\cite{cooper2023,bandstra2023}. With strategically distributed stationary radiation detectors and algorithms to fuse potential source encounters, a sparse detector network has the potential to facilitate the detection and tracking of radioactive sources moving through a city. We distinguish distributed detectors operating as an array~\cite{frigo2009}, where each detector functions independently, from detectors operating as a true detector network, enabling information exchange between detectors. The potential to improve detection capabilities by operating detectors as a network, where information from multiple detectors is aggregated and analyzed for correlations, has been studied for dense networks and gross-counting detectors~\cite{sundaresan2007,chin2010,rao2012,pahlajani2014,Flanagan2024} and spectral detectors~\cite{rao2022}. The algorithms developed to localize sources within a detector network use methods like maximum likelihood estimation~\cite{baimax,deb2013}, Bayesian methods~\cite{hite2016,bukartas2019}, particle filters~\cite{Rao2015}, and machine learning~\cite{liu2017,abdelhakim2023,Verlie2023}. The studies have shown that networking the detectors improves the overall detection performance. In this paper we focus on the binary detection/non-detection of threat sources and analyze the effectiveness of network-based detection in sparse detector configurations using physics-based simulations of the full spectral detector response to threat sources. We evaluate the performance of sparse networks with different levels of contextual data integration and investigate the ability of data fusion to enhance detection performance and study strategies with which to optimize the detector networking.
We have developed the Python Urban Deployment Model (PyUDM) simulation tool to model static detector networks in urban environments, incorporating realistic gamma-ray detector responses, radiation background models, traffic patterns, vehicle attributes, and building occlusion. After providing an overview of the simulation framework in Section~\ref{sec:simulation}, we outline a basic method for optimizing detector placement within a configuration of distributed detectors in Section~\ref{sec:placement}. A single detector is characterized in Section~\ref{sec:single_detector}, serving as a foundation for our subsequent analyses and comparisons. In Section~\ref{sec:multi_detectors}, we analyze how data fusion across an intelligent, connected network can improve detection sensitivity. We demonstrate how integrating contextual information, such as visual attributes obtained from cameras mounted on the detectors, can enhance source detection probability in a detector network. We show how the acceptable False Alarm Rate (FAR) dictates the detection threshold for individual detectors and networks of detectors.



\section{Simulation}
\label{sec:simulation}
To accurately simulate radioactive material moving through urban environments, PyUDM combines Monte Carlo simulation tools, publicly available map and traffic data, and a set of measured gamma-ray background spectra from urban environments. This section describes four key PyUDM components:
\begin{itemize}
    \item GEANT4-based detector response model for incoming gamma rays,
    \item detector background generation from measured data,
    \item traffic and map information from OpenStreetMap, Google Maps, Uber, and the City of Chicago,
    \item model for assigning vehicle attributes such as color, make, and model based on publicly available rideshare data.
\end{itemize}
PyUDM is available under a free academic license with commercial options available. Contact the corresponding author for details.
\subsection{Physics-based simulation of detector responses}
\label{detector_response}

PyUDM includes a detector response model that simulates gamma-ray signals in 2”$\times$4”$\times$16” NaI detectors. It can be expanded to include other detector shapes and materials. We build the detector response by simulating the energy deposition of incoming gamma rays using GEANT4~\cite{Geant4}. The GEANT4 model tracks event-by-event energy deposition, filling histograms in units of effective area. Histograms for varying source angle and emission energy make up a response matrix. The detector response matrix covers gamma-ray energies from 10 to 3000\,keV in steps of 50\,keV and 192~angles covering 4~pi (using HEALPix~\cite{healpix2} $\text{nside}=4$). When a simulated source-carrying vehicle drives by a detector, the detector response is sampled from the response matrix every time step of the simulation. The detector response is first interpolated to the source-detector angle and source energy, and then scaled to the source-detector distance, source activity, and time step duration. This method allows us to use the accurate but computationally expensive GEANT4 responses through computationally inexpensive interpolations rather than a full gamma-by-gamma simulation. More details on the detector response simulation and interpolation between simulated energies and angles are presented in~\cite{abgrall2024}. 

In addition to the detector, the simulated geometry may include ground and buildings that add realism to the detector response, such as down-scattered gamma rays. In this work, we chose to simulate the detector in a vacuum for simplicity. The simulated detector response covers the main expected spectral features but is missing lower-energy features from down-scattered gamma rays.



\begin{figure}[htbp]
    \centering
    \includegraphics[width=0.9\columnwidth]{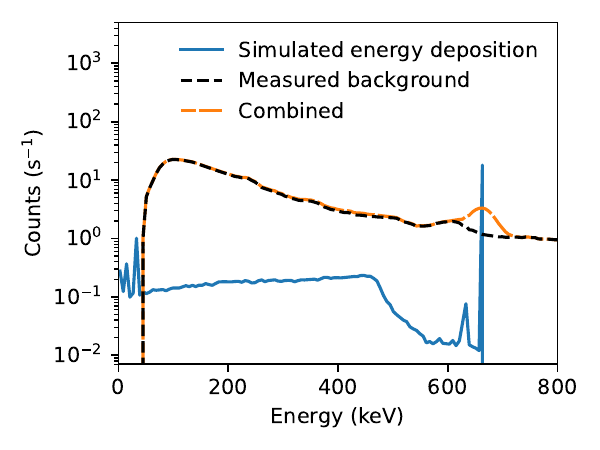}
    \caption{A sample background spectrum (dotted black), simulated $^{137}$Cs energy deposition (blue), and the combined detector response (orange, long dashes) after applying the detector resolution and hardware threshold functions. The bins are 6\,keV wide.}
    \label{fig:simspec}
\end{figure}

Source templates for a specific source of interest (e.g. $^{137}$Cs, $^{60}$Co, $^{133}$Ba) can be generated from the detector response matrix. We fetch primary gamma-ray lines and relative intensities by querying NNDC~\cite{NNDC2} and use those to build a spectrum of the primary gamma-ray energies. 
We apply blurring to the simulated detector response to mimic the resolution of measured detectors. We use an energy-dependent Gaussian blur kernel defined as $FWHM(E) = A^2 + B^2E + C^2E^2$. The parameters $A$, $B$, $C$ are fit using experimental data, and vary from detector to detector. Fig.~\ref{fig:simspec} shows the energy deposition of a simulated $^{137}$Cs source in blue, and the detector-resolution blurred response in orange.

\subsection{Measurement-based detector characteristics}
A detector's placement in a city as well as minor variations in detector electronics and scintillators result in detectors of the same model measuring different background spectra. Sampling from measured spectra is a good way to account for variability in detector components and capture the typical variation in background radiation seen in urban environments. In PyUDM, we use 1~year of measured urban radiation data from the Northern Virginia Array (NOVArray) project~\cite{NovArray2}. During a PyUDM simulation, background spectra are sampled from NOVArray data and scaled to the duration of a simulation time step. Each simulated detector inherits background characteristics, resolution, and hardware threshold from 1 of 11~NOVArray detectors. To replicate the detector resolution, a convolution matrix is calculated from fits of the NOVArray detector resolution and then applied to the simulated energy deposition described above in Section~\ref{detector_response}. The rising edges of the NOVArray detectors' background spectra are used to set the hardware thresholds in the simulated detector responses. This edge is determined to be from the first occupied channel to the peak occupied channel and is typically found around 50\,keV for NOVArray detectors. For each time step in a PyUDM simulation, background and source spectra are generated for every detector that has a source within a given radius. Background and source spectra are stored separately and when summed, they form a detector's output for that time step.

Gamma-ray background variations from radon progeny rainout and washout~\cite{bandstra2023} are not yet implemented and will be added in the next version of PyUDM.

\subsection{Traffic and map data}
PyUDM reads in map data from OpenStreetMap~\cite{OSM2}. A city can be queried by name and PyUDM will download building footprints, street layout, information about street type (motorway, road, pedestrian street etc.), number of lanes, and speed limit. Building footprints are used to calculate the occlusion of gamma rays. If there is a building between a gamma-ray source and a detector, the detector is considered occluded, and no gamma rays will be registered from that source at that position. The number of lanes on a given street is used to assign simulated vehicles a new random lane every 20 seconds of simulation time. The simulated traffic is based on ride data made public by Uber for the City of Chicago~\cite{UberTrips2}. Using the start and stop times and locations, the simulated vehicles move in the patterns provided by the rideshare data. The public rideshare data are limited to start and stop locations and do not contain the path taken. We use routes provided by Google Maps to fill in the path using their recommended drive route and expected drive time between the start and stop locations. The simulated rides presented in this paper are based on 9000~rides selected with start and/or stop locations within a $6\times6$\,km$^2$ area of downtown Chicago, covering 91\% of the public roads within selected area.
In 2006, the city of Chicago measured and published average traffic rates at 1382~positions around the city~\cite{ChicagoTraffic2}. We use the average traffic rates to scale up the number of Uber-based routes in the simulation until enough Uber-based vehicles fill the simulation to minimize the absolute difference to the measured average traffic rates. PyUDM may be extended to simulate other cities where similar traffic data is available, such as New York\cite{tlc}.

\subsection{Vehicle attribute assignment}
In addition to a radiation detector, a detector node may feature a camera running image recognition that can identify the types of vehicles passing to aid in detecting radiation anomalies. In this paper we focus on cars and model the distribution of simulated vehicles on the publicly available information of makes, models, and colors of cars provided by rideshare companies for the city of Chicago~\cite{UberVehicles2}. The information is used in PyUDM to assign a make, model, and color to each simulated vehicle, which are currently limited to cars.

\section{Detector placement\label{sec:placement}}
\begin{figure}[htbp]
    \centering
    \includegraphics[width=0.95\columnwidth]{"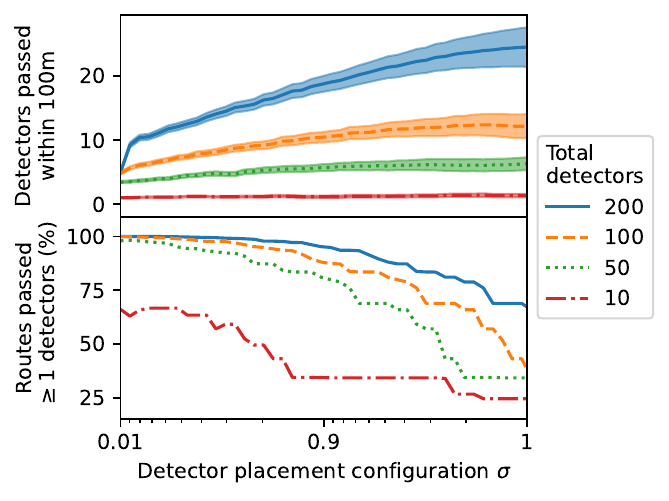"}
    \caption{Detector encounters for different network size and placement priority parameter, $\sigma$, over a $6\times6$\,km$^2$ area of downtown Chicago. The top plot shows the average number of detectors that the simulated vehicles passed within 100\,m. The shaded bands indicate $\pm0.1$ standard deviations. The bottom plot shows the percentage of the 9,000 simulated vehicles that passed at least one detector within 100\,m.}
    \label{fig:detectors_passed}
\end{figure}

Fig.~\ref{fig:detectors_passed} shows how denser detector placements increase the probability of multi-detector encounters for a vehicle driving through the $6\times6$\,km$^2$ area of downtown Chicago. The more detectors a vehicle passes, the higher the chance of detection. Conversely, routes that do not intersect with any detector nodes will not trigger an alarm, no matter how advanced an algorithm is used to analyze the detector data.
Detector placement is crucial for the performance of an urban detector network~\cite{Liu2011}. A detector at a low-traffic intersection contributes less data on passing vehicles compared to one at a busy location. If vehicles of interest follow typical traffic patterns, a detector at a busy location is more likely to detect them. However, this assumption may not always hold true. We placed detectors using the strategy outlined in Algorithm~\ref{alg:detector_placement}. This algorithm scores potential locations based on how many of the 9,000 prepared vehicle routes pass within 100\,m. This distance was chosen based on the simulated results that will be shown in Section~\ref{sec:single_detector} and Fig.~\ref{fig:SNR_heatmap}. The score is calculated to prioritize new detector locations, intercepting routes not covered by existing detectors. A detector placement priority hyperparameter, $\sigma \in {0,1}$, balances the reward between maximizing spatial coverage (the fraction of vehicle routes seen within 100\,m, $\sigma = 0$) and maximizing traffic exposure (total times vehicles are seen, $\sigma = 1$).Rewards for covering routes that have also been covered by previously placed detectors are decreased. With an $\sigma = 0.9$, a detector location that adds a fifth view of a route is rewarded only half the score ($0.9^5\approx0.5$ point) compared to a location that adds coverage of a previously unseen route (1 point). A low $\sigma$ results in a higher area coverage, with more of the vehicle routes passing at least one detector. A high $\sigma$ places more detectors where more routes pass, resulting in more routes passing multiple detectors, but sacrifices coverage on streets with lower traffic. The optimal value of $\sigma$ for a detector network to balance between prioritizing coverage or exposure will depend on the threat scenario.

\begin{algorithm}[htbp]
\caption{Place 200 detectors}\label{alg:detector_placement}
\begin{algorithmic}
\State \textbf{Input:} Generated set of routes $routes$, placement parameter $\sigma \in [0, 1]$
\While{$|{detector\_locations}| < 200$}
    \For{$route$ in $routes$}
        \State $n \gets \texttt{number\_detectors\_passed}(route,$
        \hspace*{45pt}$detector\_locations)$
        \State $route\_values[route] \gets \sigma^{n}$
    \EndFor
    \State $location\_scores \gets \texttt{score}\_\texttt{locations}(routes,$
    \hspace*{30pt}$route\_values)$
    \State $detector\_locations\texttt{.append}($
    \hspace*{30pt}$\texttt{highest\_score\_location}(location\_scores))$
\EndWhile
\end{algorithmic}
\end{algorithm}

A vehicle route is marked as passing by a detector if it is within 100\,m of the detector at any point. Algorithm~\ref{alg:detector_placement} places the first detector node at the location with highest traffic based on the generated routes. Of all the generated routes, $\sim$12\% pass within 100\,m of this location. We placed up to 200~detectors within the same $6\times6$\,km$^2$ area of downtown Chicago using different placement priorities, $\sigma$. The number of detectors seen on average by a vehicle is between 0~-~20, as shown in the top plot of Fig.~\ref{fig:detectors_passed}. The average increases with number of detectors placed, naturally, and also with $\sigma$. The bottom plot shows the fraction of all routes that are seen by at least one detector. Lower $\sigma$ prioritize placing detectors at locations that catch unseen routes at the cost of having lower numbers of detectors seen per route on average.


\begin{figure*}[htbp]
    \centering
    \includegraphics[width=0.95\textwidth]{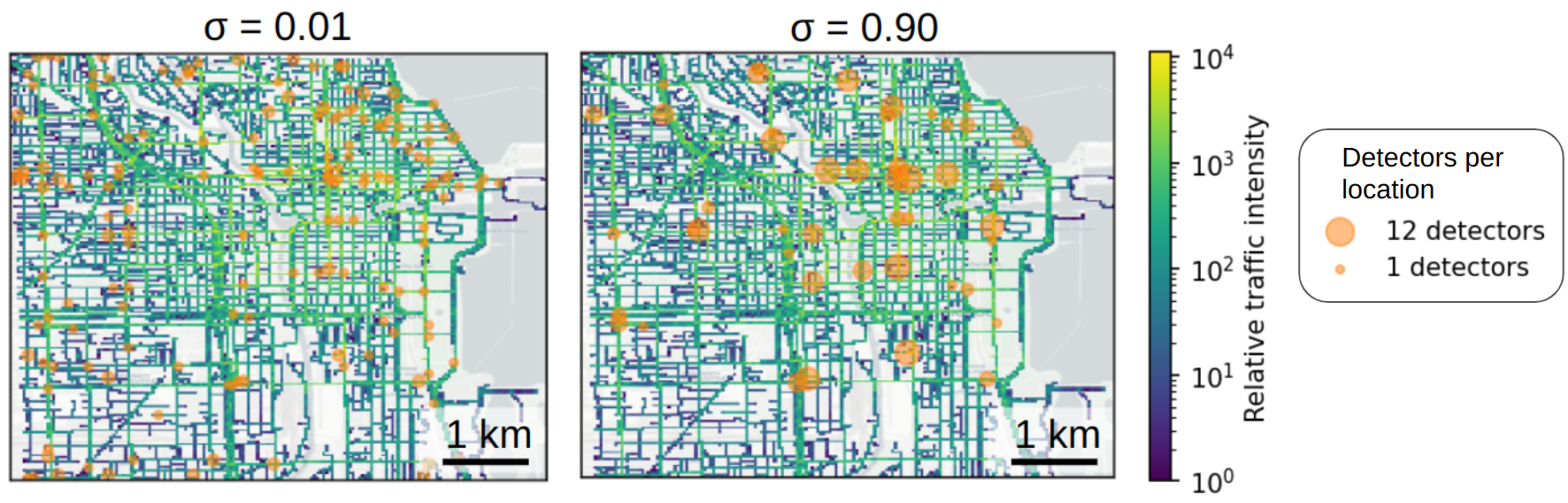}
    \caption{Two alternate configurations of 200~detectors placed over a $6\times6$\,km$^2$ area of downtown Chicago. The left map prioritizes seeing as many vehicles as possible at least once and the right map prioritizes the number of vehicle encounters, including seeing the same vehicle multiple times. The street network is colorized by relative traffic intensity.}
    \label{fig:detector_placement}
\end{figure*}

Fig.~\ref{fig:detector_placement} shows the detector node locations chosen by the placement algorithm with priority of spatial coverage (left) and traffic exposure (right). By placing 200~detectors more evenly distributed with $\sigma=0.01$, 100\% of the routes will see at least one detector and the average route passes $\sim$5~detectors. By grouping the detectors at busier intersections with $\sigma=0.90$, only $\sim$80\% of all routes are seen but the average route will encounter $\sim$5~times as many detectors as in the $\sigma=0.01$ configuration.


\section{Single detector vehicle pass-by\label{sec:single_detector}}
In this section we describe how we determine whether a measured spectrum from a source-carrying vehicle is significantly anomalous to be detected. Let us begin by looking at the relatively simple scenario of a source-carrying vehicle driving past a single 2”$\times$4”$\times$16” NaI detector on a straight road. The number of gamma rays registered in the simulated detector depends on the activity of the source, the speed of the vehicle, and the closest approach between source and detector. Background signal in the detector is scaled and Poisson-sampled from templates at each simulation time step. We calculate the signal-to-noise ratio (SNR) within an isotope-specific region of interest of a spectrum as

\begin{equation}
    SNR = \frac{\text{Source counts}}{\sqrt{\text{Source} + \text{background counts}}}.
\end{equation}

The simulated SNRs correspond to how difficult it is to discriminate source signal from background. To predict the presence or absence of a source in a spectrum, we employ an anomaly detection algorithm based on Non-negative Matrix Factorization (NMF)~\cite{nmf2019}. The method has been shown to be an effective way of using spectral template matching to detect anomalies among background spectra. The algorithm is given a long (1000\,s) background spectrum to compare incoming spectra against. It outputs an alarm metric, representing how likely a given spectrum is to be anomalous to background, here equivalent to containing a source. Spectra resulting in alarm metrics above a set threshold are predicted to contain a source, and those below the threshold are not. The sensitivity of an anomaly detection algorithm, as well as the rate of false positives, will increase with lower thresholds. In this analysis, we set the threshold to result in a False Alarm Rate (FAR) of 1~in 8~hours. Since every alarm will require attention from personnel in a deployed system, a FAR of 1~in 8~hours can be suggested as an acceptable rate and corresponds to an SNR threshold of 2.5 for the NMF-based anomaly detection~\cite{Aucott2014}. If more detectors are reporting alarms in parallel, the threshold will need to be increased to maintain the target overall FAR.

\begin{figure}[htbp]
  \centering
  \includegraphics[width=0.85\columnwidth]{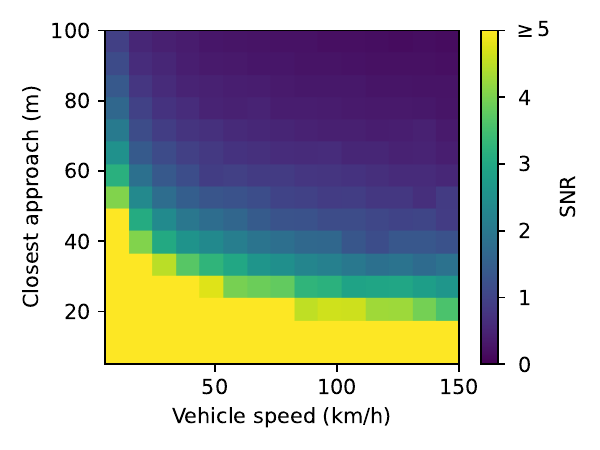}
  \caption{Simulated SNRs from an unshielded, 1000\,µCi $^{137}$Cs source passing a 2”$\times$4”$\times$16” NaI detector in a straight line.}
  \label{fig:SNR_heatmap}
\end{figure}

As the intensity of gamma rays from the source reaching the solid angle subtended by the detector falls off with the inverse square of the distance, the optimal integration time for the detector is centered around the closest point of approach assuming constant vehicle speed. To find the detector integration time that results in the highest SNR, we calculate the SNR for cumulative sums of source and background signals in 0.1\,s steps until a peak is found. We calculated the highest achievable SNR (averaged from 10 pass-bys) for car speeds between 10~-~150\,km/h and closest approaches between 5~-~100\,m for sources of $^{137}$Cs between 50~-~5000\,µCi. Fig.~\ref{fig:SNR_heatmap} shows a heatmap of simulated SNRs for a 1000\,µCi $^{137}$Cs source.

The dark blue region (SNR~$<$~1.5) contains cases where not much source signal is measured and detection will be hard to impossible at this particular FAR. The green middle region (SNR~$\sim$2.5) is close to the detection threshold. Events in this region may be correctly identified but have a high risk of going unnoticed by the single detector. In the yellow region (SNR~$>$~4), every source leaves a clear signal in the detector making detection trivial. The middle region is of particular interest when characterizing the potential benefits of networking detectors. For these events,  a single detector may not be sufficient to confidently detect the source. However, this is precisely where networked detection and data fusion from multiple detectors can add significant benefit. By combining the data from neighboring detectors that view the source-carrying vehicle at different times and locations, the combined information could enable reliable detection of sources that would be missed by any single detector alone. The next section focuses on this concept of fusing detector data from a source-carrying vehicle passing by multiple detectors.

\begin{figure}[htbp]
  \centering
  \includegraphics[width=0.85\columnwidth]{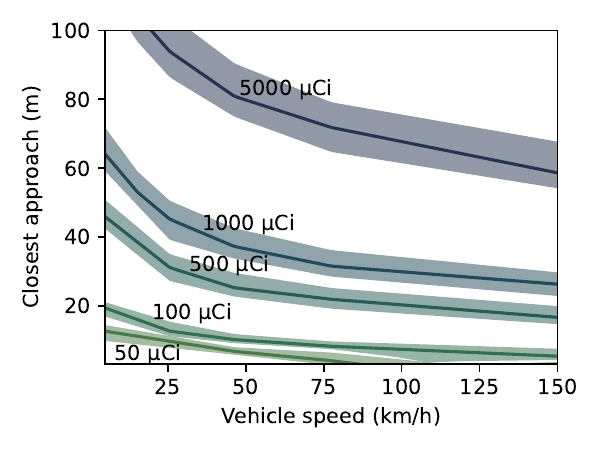}
  \caption{Bands showing the detection limit for unshielded $^{137}$Cs sources as they pass by a 2”$\times$4”$\times$16” NaI detector node at varying speed and closest approach. The bands trace the line where the SNR is 2.5, corresponding to a detection threshold with a FAR of 1~in 8~hours. The shaded areas cover SNR 2.5$\pm$0.5.}
  \label{fig:SNR_bands}
\end{figure}

In Fig.~\ref{fig:SNR_bands}, the regions of SNR 2.5\,$\pm$\,0.5 is shown for 5~different strengths of $^{137}$Cs sources. The strong 5000\,µCi source, is easily detected from more than 50\,m away, even at high speeds, while the weaker sources must pass by closer than 20\,m to the detector in order to be detected in a single pass. The SNR, and therefore the detection probability, is more influenced by the distance of closest approach than by the pass-by speed.

\section{Multi-detector deployments\label{sec:multi_detectors}}
We have simulated the response of detectors distributed over a $6\times6$\,km$^2$ area of downtown Chicago. We placed the detectors using Algorithm~\ref{alg:detector_placement}, with two different configurations. For both configurations we look at the responses when using a single detector up to 200~detectors. We let 9,000 simulated vehicles carrying $^{137}$Cs sources of strengths $5-1000$\,\textmu Ci drive through the area following typical driving patterns as described in Section~\ref{sec:simulation}. $^{137}$Cs was chosen as it is a commonly used industrial source and is a potential component of a radiological dispersal device~\cite{nemzek2004}.
We explore three methods of analyzing the data from a multi-detector system. One is the conventional approach to fielding multiple detectors and operating them independently like a detector array. The other two methods are variations of operating them as connected detector networks.
\subsection{Detector array}
The simplest way of analyzing the data of a city-wide, multi-detector system is to look at each detector one-by-one and perform anomaly detection like in Section~\ref{sec:single_detector}. If any one detector alarms in the presence of an illicit source, the source is successfully detected. This works well for source encounters that produce significant signals in a single detector~---~easier encounter dynamics below the bands shown in Fig.~\ref{fig:SNR_bands}. An array of detectors like this has an increased overall probability of detection compared to a single detector by an increased geographical coverage but the lowest SNR that can be detected is not higher than for a single detector. The detectable SNR may even be lower in practice, as the threshold for detection must be raised to maintain an overall FAR (e.g. 1~in~8~hours) when there are more detectors. For instance, 10~detectors in an array must maintain individual FARs of 1~in~80~hours for the combined array FAR to be 1~in~8~hours.


\subsection{Detector network utilizing camera information}
\begin{figure}[htbp]
  \centering
  $\vcenter{\hbox{\includegraphics[width=0.32\columnwidth]{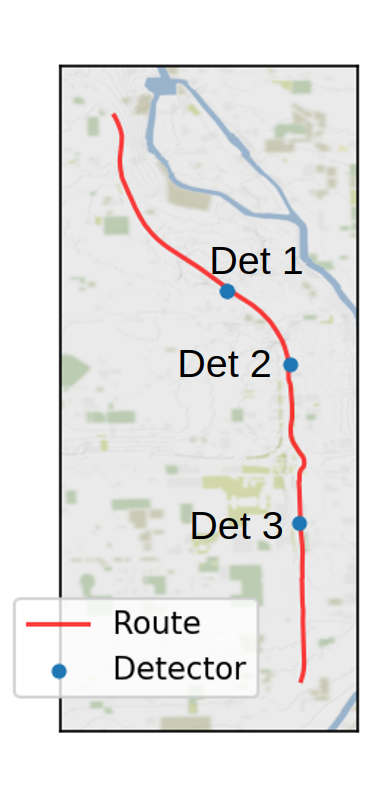}}}$
  \hspace{0.1\columnwidth}
  $\vcenter{\hbox{\includegraphics[width=0.37\columnwidth]{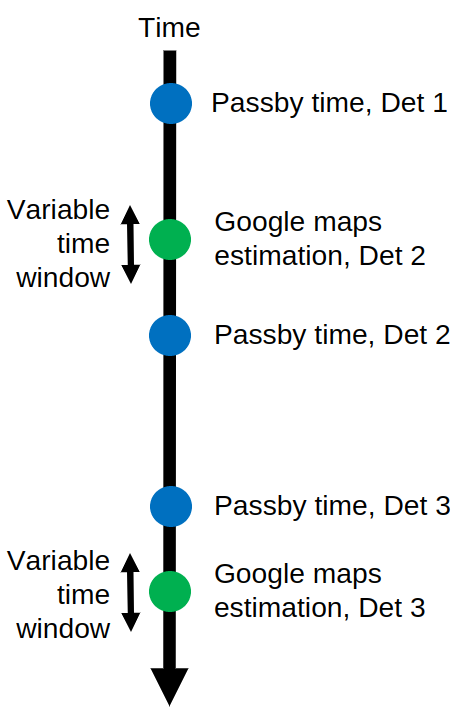}}}$
  \caption{The timeline of a source-carrying vehicle driving south, passing three detectors. The Google Maps estimated travel times are queried at the time of passing Detector 1 from the location of Detector~1 to Detector~2 and 3. The variable time windows used to consider correlated pass-bys are indicated by double-sided arrows.}
  \label{fig:timeline}
\end{figure}

For more difficult encounter dynamics~---~such as low source activity, shielded sources, greater distance between the vehicle and detector, or higher vehicle speed~---~the measured spectrum may fall below the detection threshold for a single detector, as indicated by the bands in Fig.~\ref{fig:SNR_bands}. To address this issue, it is possible to combine multiple below-threshold encounters of a source by several detectors, at different times, to achieve a statistically significant view of the source. Fig.~\ref{fig:timeline} depicts the route taken by a simulated vehicle, along with a timeline illustrating the three times this vehicle passed a detector.
Simply adding all the measured data from the three detectors taken during the duration of the vehicle drive would capture all of the source-related counts, but it would also include excessive background measured during the same time period. If the time of the vehicle passing each detector was known, one could achieve a better result by selectively adding only the optimal integration times around each detector. These times are not known in practice but can be estimated. Google Maps is able to provide reasonably accurate travel time estimates between any given locations. The travel time predictions can be employed to restrict the time frame under consideration for correlated pass-bys. We use the time of a suspected source pass-by at Detector~1 as the starting point and query Google Maps for the travel time to each neighboring detector. We can use the Google Maps estimates as the starting points in time to search for potential correlated pass-bys of the neighboring detectors. We systematically expand the time windows around the estimated time of arrival to incorporate vehicles that traveled faster and slower than the prediction, as shown by the arrows in Fig.~\ref{fig:timeline}.

\begin{figure}[htbp]
    \centering
    \includegraphics[width=1\columnwidth]{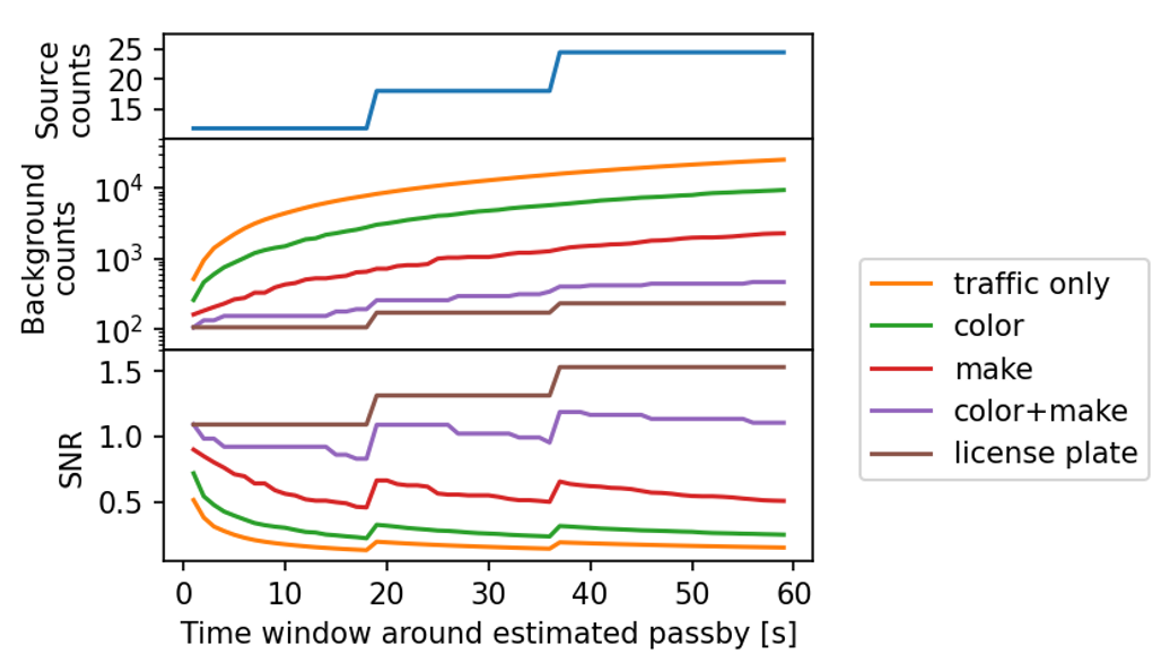}
    \caption{Source and background counts and SNR registered by a 200~detector network of which three detectors saw this source-carrying vehicle.}
    \label{fig:SNR}
\end{figure}

Fig.~\ref{fig:SNR} shows the source counts, background counts, and SNR for the source-carrying vehicle from Fig.\ref{fig:timeline}. As the variable time-window grows wider, the source counts included in the overall networked view of the vehicle also increase. Similarly, background counts also increase as the time-window is expanded. By incorporating visual data gathered at the detector nodes, further filtering of uncorrelated data can be done. If the suspected source signal occurred at Detector~1 when only a red vehicle was in sight, we can occlude any data from the neighboring nodes that does not have a red vehicle in view. Recent work has shown that it is possible to attribute signal to vehicles more precisely by combining radiation data with camera and lidar data~\cite{marshall2021, marshall2023}. As shown by the reduced background rates and increased SNR, the more uniquely a vehicle can be identified, the more accurately true correlated pass-bys can be identified. Using only the traffic information from Google Maps, the added signal from the encounters at Detectors~2 and 3 gets washed out by added background signal. Filtering for vehicles with matching color gets rid of $\sim$50\% of the uncorrelated background. If the cameras at the detector nodes are able to identify both color and make of the vehicle, the SNR increases significantly. If license plates can be read, the network has access to a unique identifier and there is no limit to the time window that can be searched in to find additional views between different detectors and times.


\subsection{Detector network finding SNR-optimal path}
An important metric for a detector network is the false negative rate; the risk of an illicit source moving through the network undetected should be minimized. In Algorithm~\ref{alg:SNR-optimal} we suggest a networked detector anomaly detection algorithm that focuses on the false negative rate by finding the single combination of detector measurement times that, when combined, would yield the highest possible SNR. We find the route that results in the highest SNR by adding 1\,s time-slices from all detectors sorted by SNR, until a peak SNR is found. The detector locations and times are compared, allowing for travel between the nodes of no faster than 3~times the speed limits. When the SNR-optimal route is identified, an anomaly test is performed. Here we simply compare the SNR to a detection threshold. The key feature of this method is: if the SNR-optimal route does not trigger an alarm, there is no other combination of spectra that could. The sensitivity to detecting a source presence is higher than the camera-based methods in the previous section as we don't have to consider all the detectors of the network and the inherent background radiation they contribute with. The increased sensitivity comes at the cost of an increased risk of overestimating source activities. 

\begin{algorithm}[htbp]
\caption{SNR-optimal anomaly test}\label{alg:SNR-optimal}
\begin{algorithmic}
\State $spectra \gets \text{read all detectors from the last hour}$
\State $sorted\_spectra \gets \text{sort $spectra$ by high-to-low SNR}$
\State $optimal\_spectrum \gets \text{initialize empty spectrum}$
\State $visited\_locations \gets \text{initialize empty list of time-stamped}$
\hspace*{0pt}$\text{locations}$
\For{$spectrum, detector\_location$ in $sorted\_spectra$}
    \If{$\texttt{snr}(optimal\_spectrum+spectrum)>\texttt{snr}(optimal\_spectrum)$}
        \If{$\texttt{possible\_drive}(detector\_location,$
        \hspace*{30pt}$visited\_locations)$}
            \State $optimal\_spectrum \gets \text{add } spectrum$
            \State $visited\_locations \gets \text{add } detector\_location$
        \EndIf
    \EndIf
\EndFor
\State\textbf{if} $\texttt{snr}(optimal\_spectrum) > detection\_threshold$
    \State \ \ \ \ \ \ \Return $alarm$
\State \textbf{else}
    \State \ \ \ \ \ \ \Return $not\_alarm$
\end{algorithmic}
\end{algorithm}

Note that we sort the spectra by SNR for computational efficiency, which is possible because we have access to the underlying source and background components of the simulated spectra. A real-life implementation of this algorithm would instead search for the combination of most anomalous spectra using, for instance, the NMF-based anomaly detection algorithm described in Section~\ref{sec:single_detector} to find the combination of times that result in the highest alarm metric and perform similarly.

\subsection{Detector Network performance\label{sec:net_res}}
\begin{figure}[htbp]
    \centering
    \includegraphics[width=0.85\columnwidth]{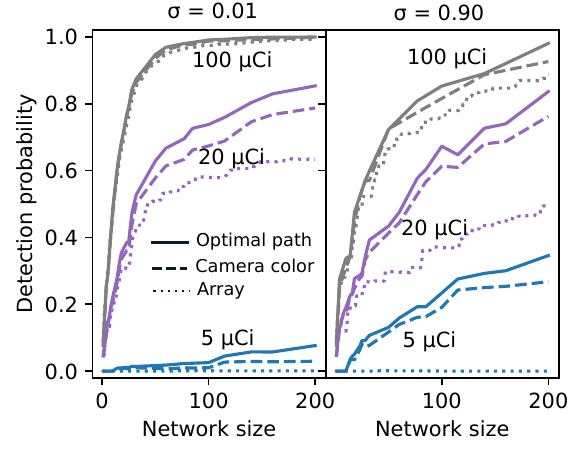}
    \caption{Probability of detecting a $5-100$\,\textmu Ci $^{137}$Cs source with detector networks/arrays of increasing size and different layouts.}
    \label{fig:TDP_strengths}
\end{figure}

Using the three methods described in Section~\ref{sec:multi_detectors} and simulated vehicles carrying $^{137}$Cs sources, we map the detection probability for the two configurations of 1~-~200~detectors from Fig.~\ref{fig:detector_placement}.The detection probability is defined as the fraction of the 9000~vehicles passing through the $6\times6$\,km$^2$ area of downtown Chicago that triggers an alarm while maintaining a FAR of 1~in~8~hours for the whole network. Fig.~\ref{fig:TDP_strengths} shows the results for the detector layout with prioritized coverage ($\sigma=0.01$) on the left and the layout with prioritized exposure ($\sigma=0.90$) on the right. The total detection probability increases with the number of detectors for all three methods, reaching close to 100\% for 100\,µCi sources for the $\sigma=0.01$ layout with 100~detectors, while the $\sigma=0.90$ layout only detects $\sim85$\% of the sources with the optimal path method and only 76\% when operated as an array. The difference between the methods of operation is bigger for the weaker sources. Operating either network as an array fails to identify any 5\,µCi sources. The networking methods manage to detect up to 35\% of the 5\,µCi sources for the $\sigma=0.90$ layout.
Optimizing detector layout for exposure is more important for weaker sources and a detector layout optimizing for coverage performs better for stronger sources.

\section{Conclusion}
\label{sec:discussion}



We have presented an algorithm for optimizing detector placement within sparse static detector networks that can be tuned via a hyperparameter to focus on maximal coverage or exposure to traffic. For a fixed number of detector nodes, node placement can be optimized to cover a larger area and can perform well at detecting stronger sources, or to focus on high-traffic areas and be more sensitive to low activity sources. 
The presented simulation tool PyUDM was used to compare three operational modes for sparse detector configurations: as an array, networked with traffic- and camera-informed search, and networked with a signal-searching algorithm. Operating detectors in a network mode, particularly with appropriate time-slicing to reduce background radiation, enhances detection capabilities, especially to weaker sources while maintaining a fixed, low FAR. Utilizing visual attributes from node-mounted cameras, such as the color and model of source-carrying vehicles, significantly narrows search time windows. This approach allows for the detection of lower activity sources with the same number of detectors or enables a sparser network to achieve the detection probability of a denser camera-less network. We introduced an SNR-optimizing network algorithm that exhibits high sensitivity to anomalous spectra but may overestimate source activities.
Achieving enhanced detection requires the ability to fuse data at the network level and benefits from leveraging in-situ contextual data. The PANDAWN network exemplifies a system that fulfills this requirement and leverages this benefit, providing an effective solution for detecting moving radioactive sources in urban environments.

These findings highlight the critical importance of networking and optimized placement strategies in improving detection sensitivity and efficiency in sparse urban detector networks.
Future work includes building on these results and implementing the suggested operational modes as deployable network algorithms and compare with existing alternatives.


\bibliographystyle{myunsrt}
\bibliography{export,odd_ones}


\end{document}